\documentclass{elsart3p}
\usepackage{epsfig}
\usepackage{color}
\newcommand{\PGRcomm}[1]{{\color{black} #1}}
\newcommand{\escomm}[1]{{\color{black} #1}}
\newcommand{\MDcomm}[1]{{\color{black} #1}}

\begin{document}

\begin{frontmatter}

\title{Deposition dynamics of Na monomers and dimers on an Ar(001) substrate}

\author{P.~M.~Dinh\corauthref{cor}$^a$}
\author{, F.~Fehrer$^b$, P.-G.~Reinhard$^b$, E.~Suraud$^a$}

\corauth[cor]{Corresponding author\\{\it Email-address}~:
  dinh@irsamc.ups-tlse.fr} 
\address{$^a$Laboratoire de Physique Th\'eorique, Universit\'e Paul
  Sabatier, CNRS, F-31062 Toulouse C\'edex, France}
\address{$^b$Institut f{\"u}r Theoretische Physik, Universit{\"a}t
  Erlangen, D-91058 Erlangen, Germany}

\begin{abstract}

We study deposition dynamics of Na and Na$_2$ on an Ar substrate, both
species neutral as well as charged.
The system is modeled by a hierarchical approach describing the Na
valence electrons by time-dependent density-functional theory while Na
core, Ar atoms and their dynamical polarizability are treated by
molecular dynamics.  We explore effects of Na charge and initial
kinetic energy of the impinging Na system. We find that neutral Na is
captured into a loosely bound adsorbate state for sufficiently low
impact energy. The charged monomers are more efficiently captured and
the cation Na$^+$ even penetrates the surface layer.  For charged
dimers, we come to different final configurations depending on the
process, direct deposit of Na$_2^+$ as a whole, or sequential deposit.
In any case, charge dramatically amplifies the excitation of the
matrix, in particular at the side of the Ar dipoles. The presence of a
charge also enhances the binding to the surface and favours
accumulation of larger compounds.
\end{abstract}

\begin{keyword}
TDDFT \sep hierarchical approach \sep deposition dynamics \sep rare
gas surface 

\PACS 
31.15.ee \sep 31.70.Hq \sep 34.35.+a \sep 36.40.Wa \sep 61.46.Bc
\end{keyword}
\end{frontmatter}

\section{Introduction}

Clusters on surfaces are a much studied subject due to its interesting
perspectives for basic research and for applications to
nano-structured materials \cite{Hab94b,Bin01}.  One important aspect
is here the synthesis of deposited clusters.  Two different techniques
have been developed, namely controlled growth of elementary units on a
surface by molecular beam epitaxy (for a brief review, see e.g.
\cite{Bru00}) or direct deposition of size-selected clusters on  a
substrate (see e.g. \cite{Har00}). An interesting aspect also concerns  a
non-destructive deposition technique of metal clusters on metal
surfaces that can be achieved by means of a thin rare gas film above
the metal surface (see e.g.~\cite{Lau05} and refs. therein).
We take up this scenario and aim here at a theoretical study of
deposition of Na on a rare gas surface. Thereby we concentrate on  the
first stages of growth, the capture of atoms and molecules with
particular emphasis on charged projectiles.

The theoretical description of deposition dynamics employs
predominantly classical molecular dynamics with effective atom-atom
forces, see \cite{Xir02}.  This was done, e.g., for the deposition
dynamics of Cu clusters on metal~\cite{Che94} or Ar~\cite{Rat99}
surfaces, and of Al or Au clusters on SiO$_2$~\cite{Tak01a}. 
That, however, does
ignore possible effects from electronic degrees of freedom, as it can
become crucial in metal clusters, and the more so if a finite net charge
is involved. One then better uses models which take care of the
electronic degrees of freedom.  Fully detailed calculations have been
undertaken, e.g., for the structure of small Na clusters on
NaCl~\cite{Hak96b} or the deposit dynamics of Pd clusters on a MgO
substrate~\cite{Mos02a}. But the expense for such fully fledged
quantum simulations grows huge.  These subtle models are hardly
extendable to truly dynamical situations, to larger clusters or
substrates, and to systematic explorations for broad variations of
conditions.  There thus exists a great manifold of approximations
which aim at an affordable compromise between reliability and expense,
often called quantum-mechanical-molecular-mechanical (QM/MM)
models. They have been applied for instance to chromophores in
bio-molecules~\cite{Gre96a,Tap07}, surface
physics~\cite{Nas01a,Inn06}, materials
physics~\cite{Rub93,Kur96,Ler98,Ler00}, embedded
molecules~\cite{Sul05a} and ion channels of cell
membranes~\cite{Buc06}.  We take up here a QM/MM modeling which was
developed particularly for the combination of Na clusters with Ar
substrate \cite{Dup96,Ger04b,Feh05a}. This method has already been
successfully applied to deposition dynamics on finite Ar
clusters~\cite{Din07a} or on Ar surfaces~\cite{Din07b}. The
originality of this approach lies in the fact that the substrate
polarizability is treated dynamically, a key aspect as soon as charged
species are considered.
In this paper, as stated above, we focus on the basic initial stages,
that is, deposition of a single Na atom or a Na dimer. We study the
effect of the initial kinetic energy given to the deposited system and
of its charge. We aim at the observation of different possible energy
thresholds between regimes of dynamical bouncing, binding or inclusion
of the Na in the Ar matrix. We also compare the direct deposition of
Na$_2$ with the sequential process where the system is deposited atom
by atom, in the spirit of the technique of atomic layer epitaxy or
deposition~\cite{Sun90}.

\section{Model}
\label{sec:model}

We start with a very brief summary of the hierachical description of
the combined Na\@Ar system. We treat the metal atoms in full
microscopic detail at the level of Time Dependent Local Density
Appromixation (TDLDA) for the valence electrons, coupled to Molecular
Dynamics (MD) for the ions. Details on the successful TDLDA-MD
approach for free clusters can be found in~\cite{Cal00,Rei03a}. The
substrate consists out of Ar atoms to which we associate classical
degrees of freedom for position and dipole moment. The latter serves
to take into account the dynamical polarizability of the substrate.
The Ar atoms are coupled to the Na by long range polarization and some
short range repulsion to account for the Pauli blocking of cluster
electrons in the vicinity of the Ar cores. The model is calibrated to
measured properties of typical Na-Ar systems. We refer the reader
to~\cite{Dup96,Ger04b,Feh05b} for a detailed description of the model.

The Ar(001) surface is modeled through six layers of 8$\times$8 Ar
atoms.  The atoms in the two lowest layers are frozen at bulk crystal
positions. The layers are periodically repeated in both lateral
directions, thus simulating bulk material in these two dimensions.  The
six-layer sample in vertical direction is finite but is sufficiently
large. We have counterchecked that by repeating some calculations
for eight layers (making 512 atoms). That did not make much a
difference and the now dynamically free fifth and sixth layers did not
acquire any sizeable amount of kinetic energy. Thus freezing them in
the 384 atom sample is a good approximation, at least for qualitative
purposes. 
%

The dynamics is initialized by placing the projectile (Na atom, ion or
dimer) at a distance of 20 a$_0$ from the surface and boosting it with
a given initial kinetic energy $E_0$, towards the substrate and 
along the direction (denoted by $z$ in the following) normal to it. We
analyze the subsequent dynamics in terms  of detailed ionic and atomic
coordinates as well as of the various parts of the kinetic energy. 

\section{Dynamical deposition of neutral Na on Ar surface}

\begin{figure*}
\begin{center}
\epsfig{file=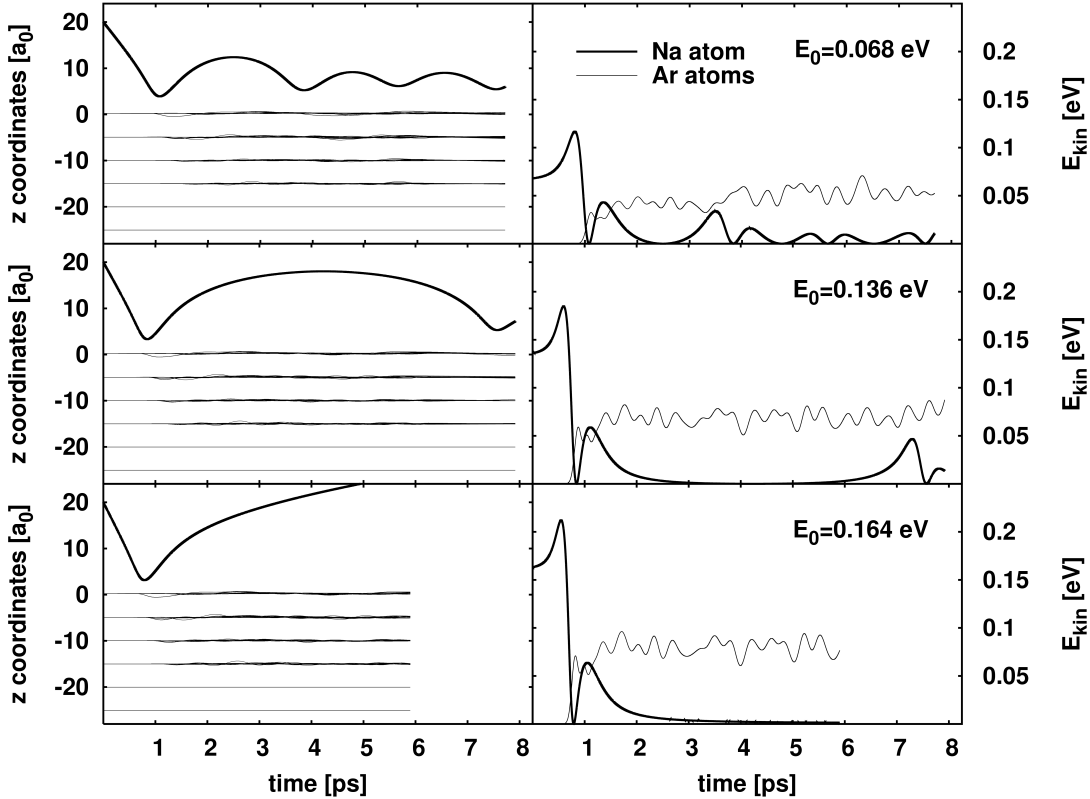,width=0.9\linewidth}
\caption{\label{fig:NaArsurf}
Time evolution of $z$ coordinates (left) and kinetic energies (right)
for the collision or the deposition of a single neutral Na atom (thick
line) on a planar Ar surface (Ar$_{384}$, faint curves), for three
different impact energies $E_0$, namely 0.218 eV (top), 0.136 eV
(middle), and 0.068 eV (bottom).}
\end{center}
\end{figure*}
%
As a first test case, we study the deposition of a neutral Na atom.
Figure \ref{fig:NaArsurf} shows results for three different impact
energies. The atom is captured by the surface for all initial kinetic
energies $E_0\leq 0.14$ eV.  Above this value, after impact, the Na
acquires a positive escaped velocity which does not change of sign
with time later on. The projectile is thus reflected and the process 
turns into an inelastic collision.  The threshold value of 
0.14 eV looks low at first glance. It is, however, already larger than
the energies of Ar binding (typically 0.05 eV) and of the NaAr dimer
(0.005 eV).
And even in the regime of capture, the atom can still have huge
amplitudes in its first bouncing oscillations reaching far away from
the surface which leaves these initial stages somewhat vulnerable
against perturbations. Safe deposit with immediate binding would
require very low impact energies. The Ar material is, in fact, rather
repelling to one single neutral Na atom. The latter is just loosely tied to
the surface and insertion inside is energetically much unfavourable.
That changes for Na clusters. Already small clusters as Na$_6$ or
Na$_8$ are tightly captured in a wide range of impact energies
\cite{Din07a,Din07b} and they are also favourably embedded deep inside
Ar material \cite{Feh07c}. The difference stands in a larger
polarizability of the cluster, due to the cooperative response of the
valence electron cloud, while one single and tightly bound electron in
the Na atom is too weak to develop a strong polarization. This
difference shows the enormous importance of the polarization
interaction in material combinations where metals and polarizable
media are involved. We will see that again when considering charged
projectiles farther below.

The reaction of the matrix seems weak when looking at the local
positions in the left panels of figure
\ref{fig:NaArsurf}. Nonetheless, one can spot a faint sound wave
propagating through the layers and some oscillations.
A more telling view of energy transport is provided by the right
panels of figure \ref{fig:NaArsurf} which show the evolution of
kinetic energies for various impact energies. The pattern are to some
extent all similar. About half of the initial kinetic energy of the Na
atom is very quickly transferred to Ar at first impact followed by a
phase where kinetic energy is flowing away from the Na at a slower
time scale. This second energy loss is due to the Na atom trying to
escape against the attractive dipole force of the surface. A small
fraction of the potential energy thus worked up is further transmitted
to the kinetic energies of the Ar atom. After 2-3 ps, we have the
typical result that almost all Na energy is transferred to the
Ar substrate which seems to share it half and half into kinetic and
potential energy.
A quick note on the lowest right panel. It looks as if the Na atom had
lost all its energy although the spatial picture (lowest left panel)
shows final reflection. At second glance, we see that a small amount
of kinetic energy remains steadily in the Na atom. It is obvious that
just above threshold, we encounter a very inelastic collision.

\section{Dynamical deposition of charged Na on Ar surface}

\subsection{The cation case}

\escomm{
As clusters are usually manipulated as cations, it is 
especially interesting to consider the deposition problem with such 
charged species. We shall now consider this case in detail. 
  
It should be noted that treating charged species 
requires a proper handling of the surface degrees of 
freedom. Accounting for the 
polarizability of Ar atoms dynamically, as we do in our model 
(see section \ref{sec:model}), is here a crucial ingredient.
It will thus also be interesting to look at how Ar dipoles respond to
a deposition. A simple measure for this effect is
the excitation energy of the dipoles which, in our model,
scales with the square of the Ar dipole amplitudes. 
\PGRcomm{When depositing} a neutral Na (cf. figure~\ref{fig:NaArsurf}),
the dipole excitation energy turns out to be vanishingly small,
whatever the deposition energy. Thus we did not show it on
the figure. Still, the effect of dipoles is known to be decisive, even
at low energies, see for example in the analysis of optical properties
of embedded metal clusters \cite{Feh05a}.
}

\subsubsection{Time evolution of positions}

\begin{figure}
\begin{center}
\epsfig{file=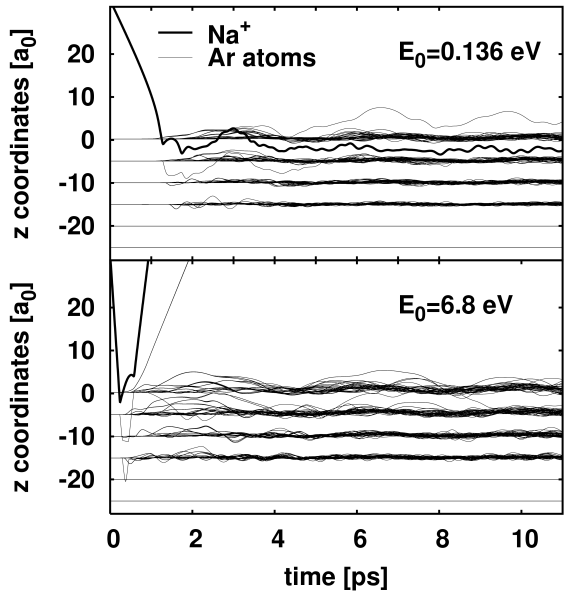,width=\linewidth}
\caption{\label{fig:napatoms}
Time evolution of $z$ coordinates from the dynamical deposition of a
Na$^+$ ion (thick lines) in Ar$_{384}$ (thin curves) with two initial
kinetic energy $E_0$ of 136 meV (top) and 6.8 eV (bottom).
}
\end{center}
\end{figure}

%
\escomm{
We first view the deposition process in real space and plot 
again the $z$ coordinates as a function of time for the deposition 
of Na$^+$ on Ar(100). 
\PGRcomm{Figure \ref{fig:napatoms} shows results for}
 two typical deposition energies, 
one below and one above deposition threshold. 
}

The lowest energy \PGRcomm{would correspond for the neutral Na atom to
a situation} just below threshold for capture (compare with middle
panel in figure \ref{fig:NaArsurf}).  The charge of the Na$^+$
enhances the attraction to the surface due to the Ar polari\-zability.
This leads in the incoming stage to a much larger acceleration of the
Na$^+$ ion towards the surface as compared with the neutral Na atom.
But the now much larger kinetic energy at contact time (about 1~ps)
does not cause immediate reflection. The large attraction enhances, in
fact, capture. The ion uses its high kinetic energy at impact to
overcome the short range repulsion of the Ar atoms and penetrates the
first layer. It is then caught, after some oscillations forth and
back, between first and second layers. In the first bounce back at
around 3~ps, it makes space by kicking one Ar atom out of its
position.  With the creation of this vacancy, some rearrangements
occur in the highest layers at a very slow time scale such that
finally the Na$^+$ resides between first and second layers while one
Ar atom is shifted out of the surface to what could be called the next
upper Ar layer.  There is however some uncertainty about the
final fate of this atom. Indeed, although the way to equilibration is
visible in the time evolution of the matrix kinetic energy (see top
panel of figure~\ref{fig:napekin}, full thick line), the matrix is
surely not thermalized yet. At 30 ps, the adatom still looks like an
atom loosely adsorbed on the surface and will very probably remain
so. But there is no guarantee that it does not finally escape by
thermal agitation. Any weak external perturbation may destabilize that
adsorbate.

\escomm{The higher energy case, presented in the bottom panel of
figure~\ref{fig:napatoms}, displays reflection. In that case, the
Na$^+$ is quickly ejected together with a few nearby Ar atoms. The
remaining surface atoms accomodate the perturbation in a way similar
to the lower energy case,} \PGRcomm{however with somewhat larger
oscillations.  }

\subsubsection{Time evolution of energies}

\begin{figure}
\begin{center}
\epsfig{file=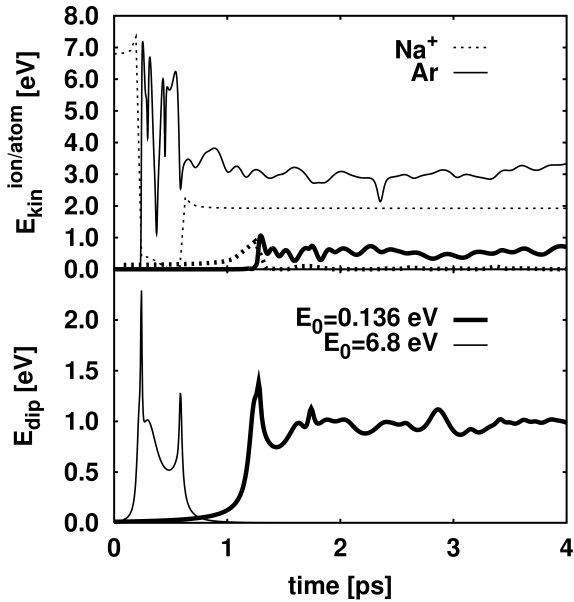,width=\linewidth}
\caption{\label{fig:napekin}
Time evolution of typical energies involved in the dynamical
deposition of a Na$^+$ ion in Ar$_{384}$, with two initial kinetic
energies $E_0$ of 136 meV (thick curves) and 6.8 eV (thin
curves). Top~: Kinetic energies of Na$^+$ (dots) and Ar atoms (full lines)
in the top panel. Bottom~: Excitation energy of Ar dipoles for
different $E_0$ as indicated.
}
\end{center}
\end{figure}
The evolution of atomic and ionic kinetic energies is plotted in the
top panel of 
figure~\ref{fig:napekin}. It again shows first a large initial
acceleration 
of the charged projectile caused by the attractive polarization
interaction with the substrate. This effect is much larger than in the
previous example with a neutral projectile.  Much similar to the
previous case is the immediate and large energy transfer to the
substrate at the time of closest impact. The figure does now also
show the kinetic energy of the Ar dipoles (see bottom panel).
 
\escomm{Let us first discuss the 
low energy case.}  
For the charged projectile, the energies of atoms and dipoles
are of the same order of magnitude~: 
For Ar atoms, about two thirds
of the ion kinetic energy at impact ($\simeq 0.9$~eV), while the Ar
dipoles get an excitation energy up to about 1 eV. The energy gathered
by the substrate comes from three sources~: {\it i)} From direct
conversion of the ion kinetic energy, {\it ii)} from release of
potential energy due to deformation and rearrangments of the matrix,
and {\it iii)} from the electrostatic influence of the inclusion of a
positive charge into the matrix. Later, the remaining Na
ion kinetic energy is quickly and almost completely taken up by the
matrix in the tight bounces between the layers. It is
remarkable to observe that the excitation of the matrix is as high for
the atoms as for the dipoles. This demonstrates that the
dynamic of dipole polarization plays a crucial role in the process and
that a theoretical description has properly to take that into account.
Omitting these degrees of freedom changes the deposition process
completely, as has been also shown in the case of Na
clusters~\cite{Din07b}. 

\escomm{The higher energy case displays another interesting
feature. At first glance, the dipole energy vanishes asymptotically,
\PGRcomm{different from} the low energy case. In fact, a closer look
at the curve shows that this energy reaches temporarily very high
values, before the few Ar atoms are emitted. It is thus likely that
the few emitted atoms were precisely the ones which had the largest
dipoles. A more detailed inspection of the} \PGRcomm{spatial
distribution of dipole
energies confirms that.
} 
%


\subsubsection{Systematics in deposition energy}

\begin{figure}
\begin{center}
\epsfig{file=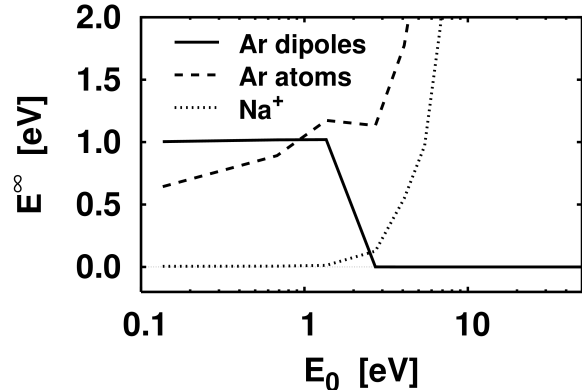,width=\linewidth}
\caption{\label{fig:napsystemat}
Asymptotic kinetic energies of Na$^+$ (dotted) and of Ar atoms
(dashes), and excitation energy of Ar dipoles (full line) as a
function of initial kinetic energy $E_0$ of the Na$^+$ projectile.
}
\end{center}
\end{figure}

\escomm{It is finally instructive to \PGRcomm{analyze the trends of}
deposition dynamics by considering the energetics
as function of the initial kinetic energy of the
projectile $E_0$. \PGRcomm{Figure~\ref{fig:napsystemat} shows} the 
asymptotic kinetic energies of Na$^+$, Ar atoms and excitation energy
of Ar dipoles as a function of $E_0$.
}
\PGRcomm{The pattern indicate a dramatic change around 
$E_0\approx 2.7$~eV, which is the transition point from capture
to reflection of the impinging Na$^+$. Above that critical
energy, the outgoing kinetic energy of the Na$^+$ increases linearly
with $E_0$. The same holds for the energy of the Ar atoms 
because it is then dominated by the few atoms which are
accompagning the departing ion.}
Below the threshold, these quantities also show a monotonous increase
with impact energy.
\PGRcomm{Most remarkable is the behaviour of the  Ar dipoles
which differs essentially from the two other quantities.
\PGRcomm{Below threshold, when} Na$^+$ ion is \PGRcomm{captured by}
the substrate, the Ar dipoles acquire an energy which seems to depend
only on the net charge and not on the \PGRcomm{initial kinetic}
energy. \PGRcomm{Above threshold, the Ar dipole energy vanishes
because} the Na$^+$ finally escapes from the surface. Comparing with
the bottom panel of figure~\ref{fig:napekin}, sharp peaks in the Ar
dipole energy appear precisely when the Na$^+$ is in the vicinity of
the surface and later on, the excitation energy rapidly decreases
towards zero as the Na$^+$ moves away.  }

A final word has to be added about the reflection threshold
we found around 2.7 eV. In the bottom panel of
figure~\ref{fig:napatoms}, we clearly notice the bouncing of the
propagating wave in the Ar substrate at the level of the fifth
layer. We recall that this layer, as well as the sixth layer, are
fixed, for reasons of computational \PGRcomm{expense. The value for
the reflection threshold, found here at $E_0 \approx 2.7$~eV, has thus
to be taken with care.  We checked the case of deposition of Na$^+$
with precisely this initial kinetic enery  on
Ar$_{512}$ (six active layers plus two fixed ones) rather than
Ar$_{384}$. Here the projectile is still captured and reflection
emerges at higher initial energies. However, the qualitative features
of deposition dynamics and trends remain the same although the
threshold value is somewhat shifted.
}

\subsection{The anion case}

\begin{figure}
\begin{center}
\epsfig{file=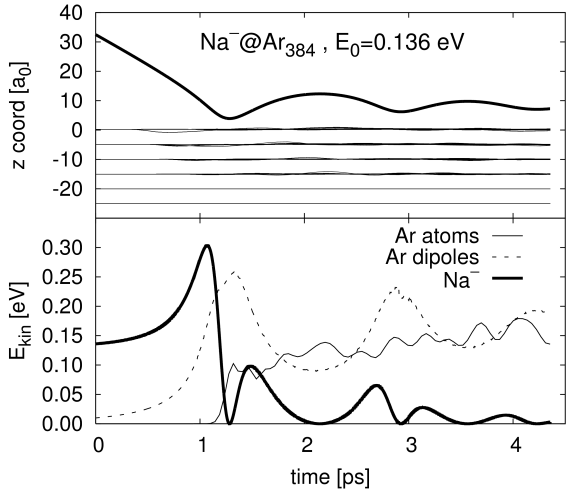,width=\linewidth}
\caption{\label{fig:NamAr384}
Dynamical deposition of Na$^-$  (thick lines) in Ar$_{384}$ (thin
curves) with an initial kinetic energy of 136 meV.
$z$ coordinates (top) and kinetic energy (bottom) of single Na and Ar
matrix as a function of time.
}
\end{center}
\end{figure}
The other choice for a charged monomer is a Na$^-$ anion.
Figure~\ref{fig:NamAr384} shows the result, this time again at the
threshold energy for the neutral case, $E_0 = 0.136$ eV. The
attraction for polarization potentials is as large as it was for the
Na$^+$ cation. But the doubly charged electron cloud experiences a
full load of the Pauli repulsion from the Ar cores which is built into
the short-range part of the effective electron-Ar interaction. As a
consequence, the Na$^-$ is blocked by the surface and will not penetrate into
the material. On the other hand, the long-range attraction
persists. Thus the anion is very efficiently captured at a safe
distance from the surface. In comparison with the neutral case, there
are no noteworthy amplitudes in the bouncing oscillations. Thus the
anion comes quickly to a rest.  In comparison to the cation case, the
impact phase is much earlier stopped such that the anion could not
acquire so much kinetic energy. This, in turn, leaves less energy to
be absorbed by the substrate. The pattern of energy transport in the
Ar atoms are similar to the neutral case~: Half of the energy is
immediately transferred at first impact and the other bits with each
bounce. However, because of the negative charge, the Ar dipoles
experience a much larger excitation and acquire a kinetic energy five
or six orders of magnitude higher than in the case of the neutral Na
deposition. Note finally that the time scale is also different because
the bounces recur much more frequently.

For both charged cases, we mention that the threshold for capture is
much higher than  for the neutral cluster (not shown). And there is
actually no regime of inelastic reflections. Enhancing the impact
energy further to force reflection leads into a regime where a
whole surface area is destroyed. This is similar to our findings for
deposition of Na clusters \cite{Din07b}.

\section{A simple analysis of the results}

\begin{figure}
\begin{center}
\epsfig{file=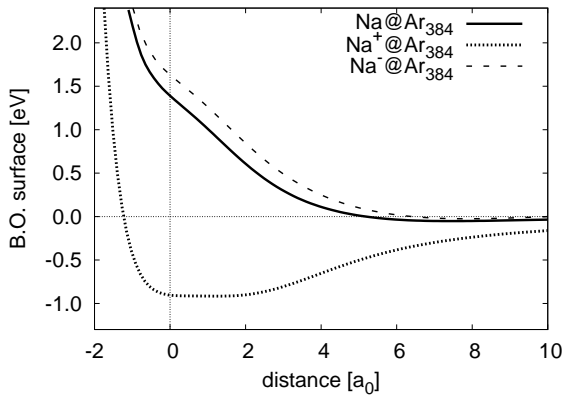,width=\linewidth}
\caption{\label{fig:bo}
Potential energy surface of neutral Na, Na$^+$ and Na$^-$ on
Ar$_{384}$, calculated with fixed Ar atoms.
}
\end{center}
\end{figure}

\PGRcomm{It is, finally, instructive to interpret the above
observed findings in terms of energy surfaces.  To that end,
we have computed the energy of Na, Na$^+$ or Na$^-$ for systematically
varied distances to the surface, keeping the distance and the atomic
positions frozen while allowing the polarizabilities to adjust to the
given configurations.  This will provide an estimate of how far or
close we are to (a)diabaticity in the deposition scenarios explored
above}.
\escomm{The energy surfaces for Na, Na$^+$ and Na$^-$ on
Ar(001)  are plotted in figure~\ref{fig:bo}. 
The ``static'' results are qualitatively compatible with}
\PGRcomm{our fully  dynamical calculations.}
Let us, for example, take the case of neutral Na. A faint minimum is
found at a distance of about 7.5 $a_0$ with a binding energy of
$-52$~meV. This has to be compared with the threshold for reflection at
136~meV that we found in our dynamical calculations. This is 2--3
times larger than the ``static'' value but the orders of magnitude are
similar.
The same qualitative conclusion holds true for Na$^+$ and Na$^-$.

\escomm{
At a quantitative level, one has to note that there remains sizable
differences between the static and dynamical results. This is an
expected and welcome feature showing that, at least in the deposition
energy range we consider, there remain genuine dynamical effects which
cannot be simply evaluated without a proper  
account of dynamics. The point was rather obvious from the behavior 
of Ar atoms and dipoles as is particularly clear from
figures~\ref{fig:napatoms} and \ref{fig:napekin}. 
}

The site dependence on the deposition has also been explored in this
``static'' way. The dynamical results presented in the previous
sections all start from a projectile initially positioned above a
hollow site in the first layer, but above on Ar atom in the second
layer. We checked that changing the deposition site of course modifies
the threshold in initial kinetic energy $E_0$ for the observation of
the reflection. This, however, does not change qualitatively our
findings. This is also compatible with the results on deposition of
Na$_6$ on Ar clusters~\cite{Din07a} and Ar surface~\cite{Din07b},
which show only weak dependence on implantation site. 


\section{Dynamical deposition of Na dimer}

\begin{figure*}
\begin{center}
\epsfig{file=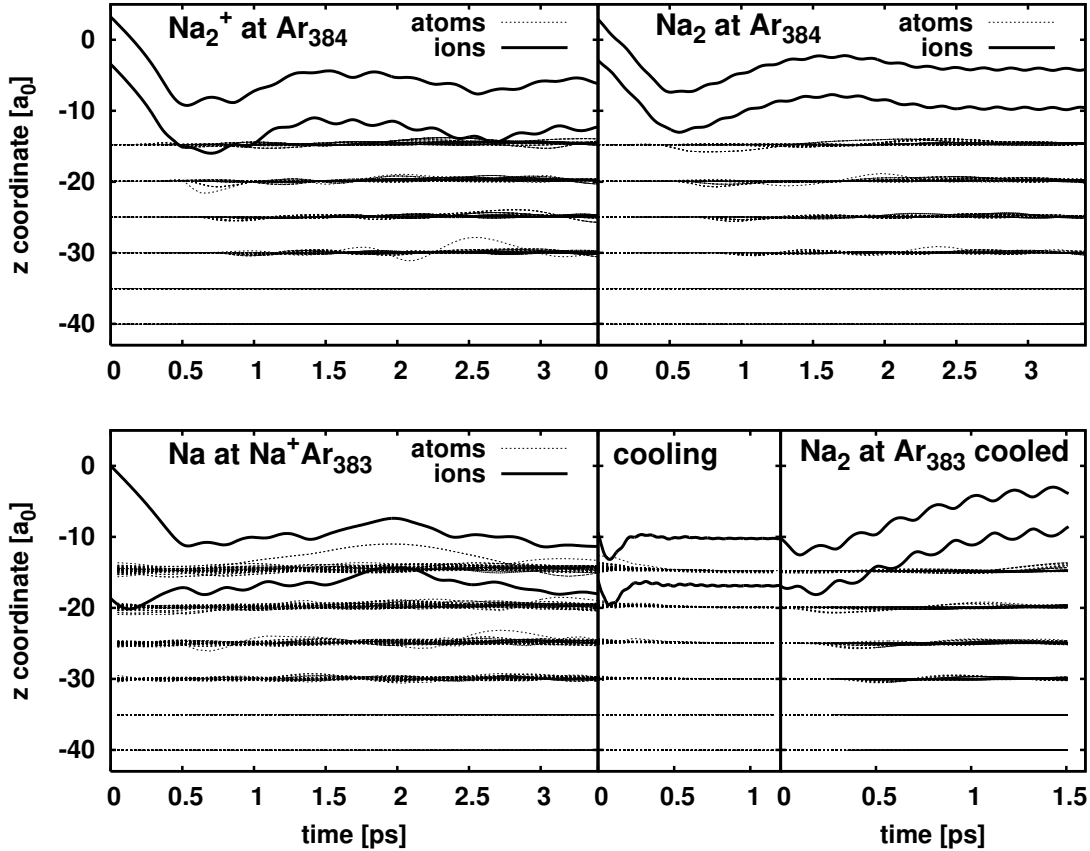,width=0.9\linewidth}
\caption{\label{fig:Na2}
Dynamical deposition of charged or neutral Na$_2$ with an initial
kinetic energy of 136 meV/ion.  Time
evolution of ionic (thick lines) and atomic (dotted lines) $z$
coordinates are shown.
Top~: direct deposition of Na$_2^+$ (left) and Na$_2$ (right) on
Ar$_{384}$. 
Bottom, sequential deposition of Na$^+$ and then of neutral Na on
Ar$_{383}$ (left) and subsequent dynamics of the obtained Na$_2^+$
which is preliminary neutralized (right).}
\end{center}
\end{figure*}

Not surprisingly, the effect of the charge is thus determinant for the
fate of the deposited atom. In the same spirit, we have compared the
cases for the dimers, Na$_2$ and Na$_2^+$. The results are shown in
the top panels of figure~\ref{fig:Na2}.
As for a single atom, the charged dimer sticks more closely to the Ar
surface than the neutral Na$_2$. Note that in both cases the dimers
are not strongly perturbed. More precisely, the dimer bond lengths
exhibit some oscillations but remain almost unchanged as compared to
the free value~: the Na$_2^+$ is slightly longer ($+2$ \%), while the
Na$_2$ bond length decreases by 4.5 \%. 
The matrix shows more perturbation
than in the previous cases with a single Na projectile. Two atoms have
simply more impact which is fully downloaded into the matrix.
Comparing the two dimers, we see that Na$_2^+$ attaches more tightly
to the Ar surface which is, again, due to the larger attraction.
However, the positively charged Na$_2^+$ stays above the surface and
does not manage to dive below as the Na$^+$ did.

There is an alternative option to bring a dimer onto the surface, that
is, sequential deposit of monomers. Particularly interesting is here the
case where a Na$^+$ cation was deposited first  and where, in a second
round, a neutral Na atom is attached.  The left lower panel of
figure~\ref{fig:Na2} shows that process.  To produce the corresponding 
initial state, we take the final state of Na$^+$ deposition (see top
panel of figure~\ref{fig:napatoms}), remove the Ar adatom, relax the
remaining  
configuration by cooling, and inject a neutral Na atom a distance of
15 a$_0$ from the surface with our meanwhile standard impact energy of
0.14 eV. The new Na atom is
captured with small remaining oscillations.  Compared with the direct
deposition of Na$_2^+$ (top left panel of same figure), the charged
dimer is more deeply bound with one leg residing below the surface. We
also observe that its bond length is almost unchanged ($-0.23$~\%)
with respect to the value of the free charged dimer. The example
thus demonstrates that the final state can depend sensitively on  the
production process. 

One could now hope that neutralization of that immersed Na$_2^+$ leads
to an equally deep bound Na$_2$ dimer. To that end, we take the final
state from the previous sequential deposit (at the end time in the
lower left panel), cool the obtained configuration, and add an
electron in the electronic ground state of the tied dimer with yet
fixed ionic configuration. Then we release the system to fully free
electronic, ionic, and atomic dynamics. The result is shown in the
right lower panel of figure~\ref{fig:Na2}. The now neutral dimer pops
up out of the surface and performs bouncing oscillations with large
amplitude about the final stage of Na$_2$\@Ar$_{383}$ which
was also obtained by direct deposit of Na$_2$, see upper right
panel. The minor difference with the now missing adatom
plays little role for the comparison. There are fast oscillations of
the dimer bond length. These emerge because the neutral dimer has a
smaller bond length than the charged one from which it was
started. Note that these bond-length oscillations persist for
long. This indicates that energy transfer from intrinsic ionic motion
to the substrate is very slow, a feature which was also observed for
larger clusters in Ar substrate \cite{Feh05b}.

\begin{figure}[htbp]
\begin{center}
\epsfig{file=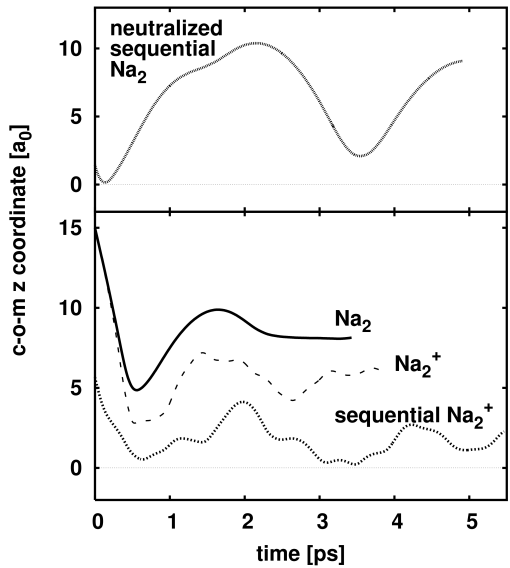,width=0.9\linewidth}
\caption{\label{fig:Na2_analysis}
Time evolution of the center of mass $z$-coordinate, relative to the
Ar surface (here at $z=0$), in a dynamical deposition of various Na
dimers, as indicated.} 
\end{center}
\end{figure}
In order to quantify the dynamics of dimer deposition in simple terms,
we have plotted in figure~\ref{fig:Na2_analysis} their center-of-mass
(c.o.m.) $z$ coordinate as a function of time.  This complements the
previous, more detailed, figures and allows a more direct comparison
between the various cases.
The lower panel shows direct deposit of neutral dimer as well as
charged dimer and sequential deposit leading to a charged dimer.  The
trend is obvious~: Binding stays 28~\% closer to the surface for the
charged dimer compared with the neutral one, 
and sequential deposit brings it even closer (of about 60~\%) so that
the dimers center lies almost at the surface.
For completeness, we also show in the upper panel the evolution of the
re-neutralized charged dimer. The return to the equilibrium position
of the directly deposited neutral dimer is visible as well as the
still large oscillations about that point. Significant energy transfer
happens only at the bouncing points and the long time span per bounce
lets us predict a very slow relaxation needing about hundreds of ps.

\section{Conclusion}
To conclude this paper, we have studied deposition of Na atoms, Na
ions and Na dimers on Ar(001) substrate, using a hierarchical
approach with time-dependent density-functional theory for the Na
electrons coupled to molecular dynamics for the Na ions and Ar atoms
as well as dipole moments.
We have paid particular attention to the effect of charged
projectiles.  We have found that the neutral Na is not likely to
penetrate into the Ar matrix and sticks loosely to the Ar surface for
initial kinetic energy lower than 0.14 eV while it is inelastically
reflected for larger energies. A Na$^+$ cation behaves much
differently. It is tightly captured and even penetrates the surface to
reside finally between surface and next layer. The Ar surface
undergoes strong pertubations and displaces one Ar atom to an adatom
site. 
\MDcomm{
At a given size of Ar substrate, we found a reflection threshold
twenty times larger for Na$^+$ than for neutral Na.}
Different is the behavior for the negatively charged Na$^-$
anion. It is also tightly bound.  But the strong electron-Ar repulsion
keeps it safely above the surface.
The deposition of Na dimers shows similar trends as for the atom.  The
charged dimer is closer bound than the neutral one. It stays, however,
fully outside the substrate.  The alternative process of sequential
deposit produces a different final state for the charged Na$_2^+$
dimer. The lower ion of the dimer is placed now below the surface
while the upper one stays just above. The final state obviously
depends on the pathway of the process. 
It was, however, not possible to keep a neutral dimer in that close
contact with the surface. Re-neutralizing the close Na$_2^+$
configuration leads back to the Na$_2$ outside the surface at a
distance which was obtained also by direct deposit of Na$_2$.
After all, the results show that charge makes a huge difference in
connection with polarizable media as, e.g., Ar substrate. It acts to
some extent as a catalyst for capture. The studies will be continued
with larger samples to explore different scenarios for producing
deposited clusters.

\bigskip

Acknowledgments: This work was supported by the DFG, project nr. RE
322/10-1, the French-German exchange program PROCOPE nr. 07523TE, the
CNRS Programme ``Mat\'eriaux'' (CPR-ISMIR), Institut Universitaire de
France, the Humboldt foundation, a Gay-Lussac price, and the French
computational facilities CalMip (Calcul en Midi-Pyr\'en\'ees), IDRIS and
CINES.


\end{document}